\documentclass[pre,twocolumn,a4paper, amsfonts, amssymb, amsmath, reprint, showkeys, nofootinbib, twoside]{revtex4-1}
\usepackage{color,graphicx,hyperref,amsmath}
\usepackage{mathtools}
\usepackage{graphicx}
\usepackage{float}
\newcommand{\im}{\mathrm{i}}

\begin{document}

\title{Influence of Asymmetric Parameters in Higher-Order Coupling With Bimodal Frequency Distribution}

\author{M. Manoranjani}

\altaffiliation{  Department of Physics, Centre for Nonlinear Science and Engineering, School of Electrical and Electronics Engineering, SASTRA Deemed University, Thanjavur 613 401, India}

\author{R. Gopal}

\altaffiliation{ Department of Physics, Centre for Nonlinear Science and Engineering, School of Electrical and Electronics Engineering, SASTRA Deemed University, Thanjavur 613 401, India}

\author{D. V. Senthilkumar}
\altaffiliation{School of Physics,
	Indian Institute of Science Education and Research,
	Thiruvananthapuram-695016, India}
\author{ V. K. Chandrasekar}
\altaffiliation{ Department of Physics,	Centre for Nonlinear Science and Engineering, School of Electrical and Electronics Engineering, SASTRA Deemed University, Thanjavur 613 401, India}
\author{M. Lakshmanan}
\altaffiliation{Department of  Nonlinear Dynamics, School of Physics,  Bharathidasan University, Tiruchirapalli - 620 024, India}

\email{chandru25nld@gmail.com}


\date{\today}

\begin{abstract}
We investigate the phase diagram of  the Sakaguchi-Kuramoto model with a higher order interaction along with the
traditional pairwise interaction. We also introduce asymmetry parameters in both the interaction terms and investigate the collective dynamics and their
transitions in the phase diagrams under  both unimodal and bimodal frequency distributions.  We deduce the evolution equations for the macroscopic order parameters and
eventually derive pitchfork and Hopf bifurcation curves.  Transition from the incoherent state to standing wave pattern is observed in the presence of
the  unimodal  frequency distribution. In contrast, a rich variety of
dynamical states such as the incoherent state, partially synchronized state-I, partially synchronized state-II, 
and standing wave patterns and transitions among them  are observed in the phase diagram, via various bifurcation scenarios including  saddle-node and homoclinic bifurcations,
in the presence of bimodal frequency distribution.
Higher order coupling enhances the spread of the bistable regions in the phase diagrams and also leads to the manifestation of bistability between
incoherent and  partially synchronized states even with unimodal frequency distribution, which is otherwise not observed with the pairwise coupling.
Further, the asymmetry parameters facilitate the onset of several bistable and multistable regions in the phase diagrams.  Very large values of the asymmetry parameters
allow the phase diagrams to admit only the monostable dynamical states.
\end{abstract}

\maketitle
Keywords: Higher-Order Coupling, Sakaguchi-Kuramoto model, Bifurcation

\section{Introduction}
\label{sec:intro}
Coupled nonlinear oscillators constitute an excellent framework to unravel and understand a plethora of intriguing collective dynamics/patterns
observed in a wide variety of natural systems~\cite{Kuramoto:1984,winfree,Strogatz:2000,Pikovsky:2001,Acebron:2005}.   In particular, the phenomenon of synchronization has been widely studied in the past two decades 
due to its manifestation in several natural and man-made systems~\cite{Pikovsky:2001,Acebron:2005,Gupta:2014,Gupta:2018}.  For instance, collective synchrony includes synchronized firing
of cardiac pacemaker cells~\cite{Peskin:1975}, synchronous emission of light pulses by groups of fireflies~\cite{Buck:1988}, chirping of crickets~\cite{Walker:1969}, 
synchronization in ensembles of electrochemical oscillators~\cite{Kiss:2002}, synchronization in human cerebral connectome~\cite{Schmidt:2015},  and
synchronous clapping of audience~\cite{Neda:2000}. 
Incredibly, the Kuramoto model has been employed as a  paradigmatic model  to 
understand diverse emerging nonlinear phenomena  across various disciplines, including physics, biology, chemistry, ecology, electrical engineering,
neuroscience, and sociology~\cite{Kuramoto:1984,winfree,Strogatz:2000,Pikovsky:2001,Acebron:2005,Pikovsky:2001}, as it allows
for an exact analytical treatment in most cases in explaining macroscopic  dynamics.

The Kuramoto model comprises  of $N$ globally-coupled phase oscillators with distributed natural frequencies interacting symmetrically with one another
through the sine of their phase differences.  Considering symmetric interaction in the dynamics is only an approximation that may
simplify the theoretical analysis, which indeed may fail to capture important phenomena occurring in real systems.  In contrast to the standard Kuramoto model, interactions between
oscillators may be asymmetric, in general.  For example, asymmetric interaction leads to novel features such as  families of traveling wave 
states~\cite{Iatsenko:2013,Petkoski:2013}, glassy states and super-relaxation~\cite{Iatsenko:2014}, and so forth, and has been invoked to
discuss coupled circadian neurons~\cite{Gu:2016}, dynamic interactions~\cite{Yang:2020,Sakaguchi:1988}, etc.  A
generalization of the Kuramoto model that accounts for asymmetric interaction is the so-called Sakaguchi-Kuramoto model, whose
 dynamics can be described by the equation of motion~\cite{Sakaguchi:1986,Oleh-SK-1,Oleh-SK-2} 
\begin{equation}
\frac{{\rm d}\theta_j}{{\rm d}t}=\omega_j+\frac{K}{N}\sum_{k=1}^N
\sin(\theta_k-\theta_j +\alpha),
\label{eq:eom-SK}
\end{equation}
where $0\le\alpha < \pi/2$ is the asymmetry parameter. The model~(\ref{eq:eom-SK}) and its variants have been successfully employed to study a variety of dynamical
scenarios such as  disordered Josephson series array~\cite{Wiesenfeld:1996}, multiplex network~\cite{sj:1,sj:2,sj:3,sj:4}, time-delayed interactions~\cite{Yeung:1999}, hierarchical populations of coupled
oscillators~\cite{Pikovsky:2008}, chaotic transients~\cite{Wolfrum:2011}, dynamics of pulse-coupled oscillators~\cite{Pazo:2014}, etc. 

Majority of the investigations  in either Kuramoto or Sakaguchi-Kuramoto models were carried out with pairwise interactions. 
Nevertheless, in many realistic systems,  such as Huygens pendulum,  
neuronal oscillators, genetic networks,  globally coupled photochemical oscillators, etc., ~\cite{o:2,o:3,o:4,o:5}
higher order Fourier harmonics  in the coupling function~\cite{ho:1,Bick2011}  or  higher order couplings~\cite{Bick2016,ho:2} 
play a predominant role in shaping the collective dynamics. 
Recently, it has been shown that higher order couplings lead to  added nonlinearity in the macroscopic system 
dynamics that induce abrupt synchronization transitions via hysteresis and bistability~{\cite{o:5}}. 
Further, higher order interactions are shown to  stabilize strongly synchronized states even when the pairwise coupling is repulsive, which is otherwise unstable~{\cite{ho:3}}.
Abrupt or explosive synchronization was shown to manifest in networks in which the degrees of the nodes are positively correlated with the
frequency of the node dynamics.  In contrast, higher oder interactions are shown to be  responsible 
for the rapid switching to synchronization, leading to explosive synchronization,  in many biological and other systems  without
the need for particular correlation mechanism between the oscillators and the topological
structure~\cite{ho:4}.

In this work, we unravel the influence of the asymmetry parameters in the phase diagram  of the Sakaguchi-Kuramoto model with  
pairwise and higher order couplings  under the influence of both the unimodal and bimodal distributions of the natural frequencies.  We  employ
two different asymmetry parameters, namely $\alpha_1$ in the pairwise coupling  and $\alpha_2$ in the  higher order coupling. 
The effect of interplay of the asymmetry parameters  and the higher order coupling on the collective dynamical behavior of the  Sakaguchi-Kuramoto model  will be captured  in
the two parameter phase diagrams.  We consider five different cases, namely
 (i)  $\alpha_1$~=~$\alpha_2$~=~0 (ii) $\alpha_1$~=~$\alpha_2~\ne~0$, (iii) $\alpha_1~\ne~0$;  $\alpha_2$~=~0,   (iv) $\alpha_2~\ne~0$; $\alpha_1$~=~0, and
 (v) $\alpha_1~>0~\&~\alpha_2>0$.
 to unravel the emerging collective dynamics and their respective  phase diagrams.  We observe incoherent state (IC), partially synchronized state-I (PS-I), 
 partially synchronized state-II (PS-II), and standing wave (SW) in the phase diagrams along with various bistable and multistable regions.  
 We also deduce the evolution equations for the macroscopic order
 parameters by employing the Ott-Antonsen ansatz~\cite{Ott:2008,Ott:2009}.  We  derive analytical stability conditions for the incoherent state, which results in
 the pitchfork and Hopf bifurcation curves, from the governing equations of  motion of the macroscopic order parameters.  Furthermore, we obtain 
 the saddle-node and homoclinic bifurcation curves using the software package XPPAUT~\cite{xpp}, which leads to several bifurcation transitions  across the various dynamical states. 
 We find that the higher order coupling essentially facilitates enlargement of bistable states.
Higher order coupling also facilitates the onset of the bistability between the IC and PS-I    states
 even for  the unimodal frequency distribution, a phenomenon which cannot be seen in the Sakaguchi-Kuramoto model with pairwise coupling and unimodal distribution. 
 Furthermore,  a low value of $\alpha_1$ for $\alpha_2=0$ and  a large value of $\alpha_2$ for $\alpha_1=0$  facilitate
 the onset of PS-II and bistable region R3 (bistability between PS-I and PS-II)
 in the phase diagram. Very large values of  $\alpha_1$  and  $\alpha_2$  allow the phase diagrams to admit only the monostable dynamical states despite the fact that appropriate 
 values of the asymmetry parameters induce bistable and multistable states. It is to be noted that bistable (multistable) regions are characterized by abrupt transitions among
 the dynamical states.
 
The paper is organized as follows.  We introduce the Sakaguchi-Kuramoto model  in Sec.~\ref{model}.  We deduce the evolution equations corresponding to
the macroscopic order parameters using the Ott-Antonsen ansatz in Sec.~\ref{sec:analysis}.  In Sec.~\ref{sec:numerics}, we illustrate the phase diagrams
of the model with both unimodal and bimodal frequency distribution for various possible combinations of the asymmetry parameters $\alpha_1$ and  $\alpha_2$ and discuss the dynamical transitions across various bifurcation scenarios
demarcating the dynamical states in the phase diagrams. Finally, we will provide a summary and conclusions in Sec.~\ref{sec:conclusions}.

\section{Model}
\label{model}
 The $N$-coupled Sakaguchi-Kuramoto model  with a specific higher order interaction is governed by the set of $N$ coupled first order nonlinear ordinary differential equations (ODEs),
\begin{align}
\label{eq:km2}
\dot{\theta}_i&\,=\omega_i+{k}\big[\frac{1}{N}\sum_{j=1}^{N}\sin(\theta_j-\theta_i-\alpha_1)\\  \nonumber
&\,+\frac{1}{N^3}\sum_{j=1}^{N}\sum_{k=1}^{N}\sum_{l=1}^{N}\sin(\theta_j+\theta_k-\theta_l-\theta_i-\alpha_2)\big],~i=1,2,\ldots, N,
\end{align}
where $\theta_i$ is the phase of  the $i$th oscillator, $\omega_i$  is its natural frequency,  which is typically assumed to be drawn from a well behaved distribution $g(\omega)$. $\alpha_1$ and $\alpha_2$ are the asymmetry parameters of pairwise and higher order interactions, respectively. $k$ is the coupling strength of both pairwise and higher order interactions~\cite{ho:1,ho:2,ho:3,ho:4}. The Kuramoto model with higher-order interactions  is known to describe  topological structures such as higher-order simplexes or a simplicial complex~\cite{ex:1,ex:2}, which are relevant to brain dynamics, neuronal networks, and biological transport networks~\cite{ex:3,ex:4}.  In recent times, neuroscience studies have  confirmed the existence of higher-order interactions between neurons. For example,  astrocytes and other glial cells are thought to be a biological source of high-order interactions since they interact with hundreds of synapses and actively regulate their activity~\cite{ex:5,ex:6}.\\

We consider a bimodal frequency distribution  for $g(\omega)$ in our system. Specifically, we  consider the Lorentzian distribution of unimodal and bimodal frequency distribution,
\begin{align}
	g(\omega)&=\frac{\gamma}{\pi((\omega-\omega_0)^2+\gamma^2)};~~\gamma >0.
	\label{eq:lor}\\
g(\omega)&=\frac{\gamma}{\pi}\left[\frac{1}{((\omega-\omega_0)^2+\gamma^2)}+\frac{1}{((\omega+\omega_0)^2+\gamma^2)}\right],~\gamma >0.
\label{eq:bil}
\end{align}
Here $\gamma$ is the width parameter (half width at half maximum)
of each peak and $\pm\omega_0$ are the location of their peaks. A more physically relevant interpretation
of $\omega_0$ is that it defines the detuning in the system (which is proportional to the
separation between the two central frequencies).
Note that the form of the distribution $g(\omega)$ given in (\ref{eq:bil}) 
is symmetric about zero. Another
point to observe is that $g(\omega)$ is bimodal if and only if the
peaks are sufficiently far apart compared to their widths.
Specifically, one needs $\omega_0 > \gamma/\sqrt{3}$. Otherwise, the distribution is unimodal and the classical results still apply.\\

\section{Evolution equation of the macroscopic order parameters}
\label{sec:analysis}
In the thermodynamic limit ($N \to \infty$ ), the system of equations  (\ref{eq:km2})  can be reduced to  a finite set of macroscopic variables
in terms of the  macroscopic order parameters governing the dynamics of the original system of equations. In this limit, the discrete set of equations 
can be  extended to a continuous formulation using the probability density function  $f(\theta,\omega,t)$,
where   $f(\theta,\omega,t){\rm d}\theta$ characterizes  the fraction of the oscillators with phases between $[\theta,\theta+{\rm d}\theta]$  along with
the natural frequency $\omega$ at a time $t$.


The distribution is
$2\pi$-periodic in $\theta$ and obeys the normalization condition
\begin{equation}
\int_0^{2\pi} {\rm d}\theta~f(\theta,\omega,t)=g(\omega)~\forall~\omega.
\label{eq:norm}
\end{equation}
The evolution of $f(\theta,\omega,t)$ follows the continuity equation
\begin{equation}
\frac{\partial f}{\partial t}+\frac{\partial(fv) }{\partial
	\theta}=0,
\label{eq:continuity-equ}
\end{equation}
where $v(\theta,\omega, t)=\frac{d\theta}{dt}$ is the angular velocity at position $\theta$ at time $t$. From Eq. (\ref{eq:km2}), one can get
\begin{align}
v(\theta,\omega,t)=\omega+&\frac{k}{2i}\big[(Ze^{-i(\theta+\alpha_1)}-Z^\star e^{i(\theta+\alpha_1)})\nonumber\\+&({Z^2} {Z^\star} e^{-i(\theta+\alpha_2)}-{Z^\star}^2 Z e^{i(\theta+\alpha_2)})\big],
\end{align}
where $Z(t)$ is the macroscopic order parameter defined as
\begin{equation}
Z=\int_{-\infty}^{\infty} g(\omega) \int_0^{2\pi}   f(\theta, \omega, t)e^{i\theta}d\theta d\omega,
\label{eq:mo}	
\end{equation}
and  $Z^\star$ is its complex conjugate.
Expanding $f(\theta, \omega, t)$ in Fourier series, we have
\begin{equation}
f(\theta,\omega,t)=\frac{g(\omega)}{2\pi}\left[1+\sum_{n=1}^\infty
\left(a_n(\omega,t) e^{\im n\theta}\right)+{\rm c.c.}\right],
\label{eq:f-Fourier}
\end{equation}
where the prefactor of $g(\omega)$ ensures that the normalization
(\ref{eq:norm}) is satisfied, $a_n(\omega,t)$ is the $n$-th Fourier
coefficient, while c.c. denotes the complex conjugation
of the preceding sum within the brackets. Using  the  Ott-Antonsen ansatz ~\cite{Ott:2008,Ott:2009}
\begin{equation}
a_n(\omega,t)=\left[a(\omega,t)\right]^n,
\label{eq:OA}
\end{equation}
one can obtain,
\begin{equation}
\frac{\partial a}{\partial t}+i\omega a+\frac{k}{2}\big[(Z {a^2}e^{-i\alpha_1}-Z^\star e^{i\alpha_1})+{|Z|^2}( {Z} {a^2} e^{-i\alpha_2}-{Z^\star} e^{i\alpha_2})\big],
\label{eq:12}
\end{equation}
where
\begin{equation}
Z=\int_{-\infty}^{\infty}a^\star(t,\omega)g(\omega)d\omega.
\label{eq:13}	
\end{equation}
\begin{figure*}[ht]
	\hspace*{-1cm}
	\includegraphics[width=18.5cm]{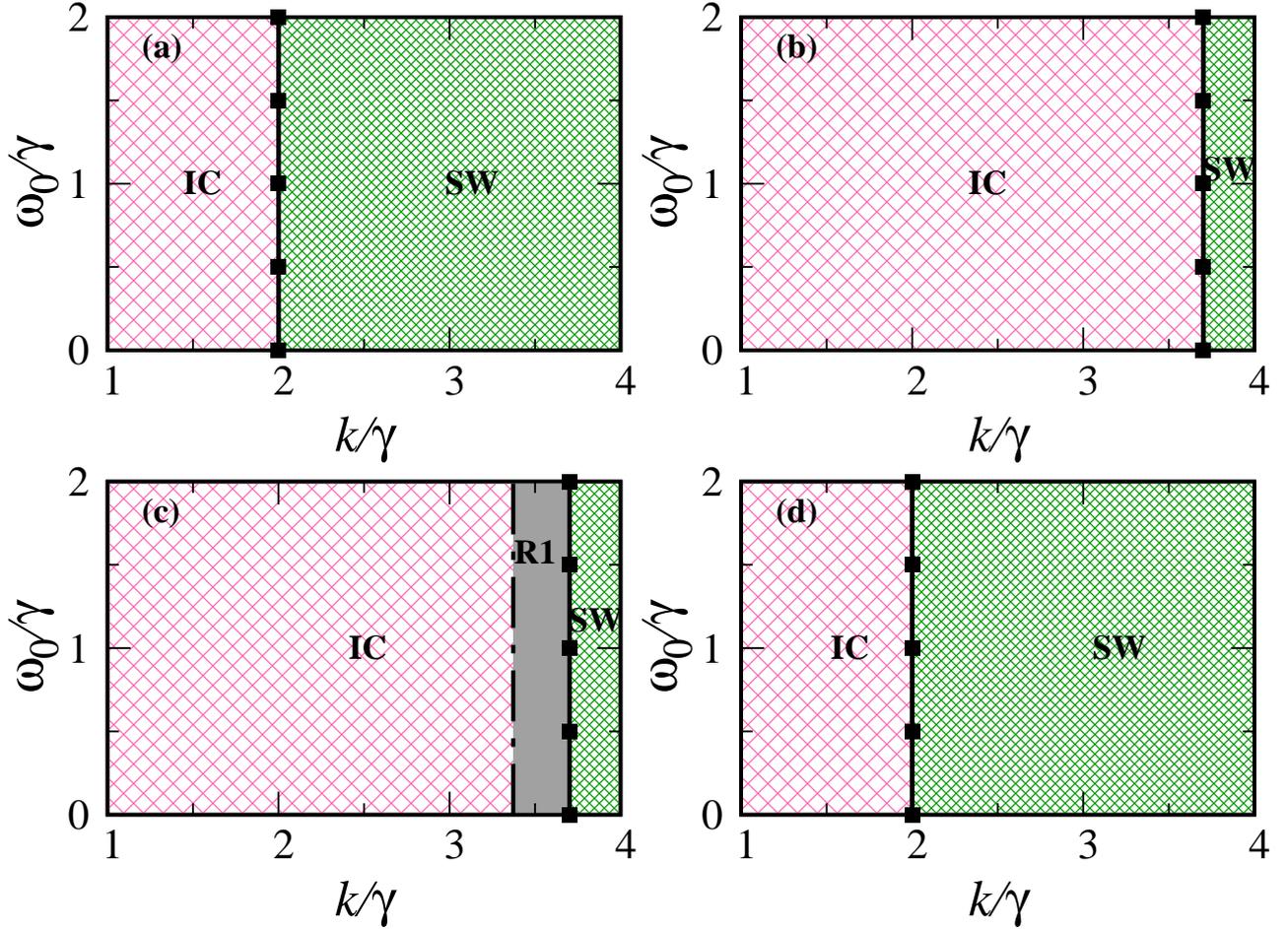}
	\caption{Phase diagrams in  the ($k/\gamma-\omega_0/\gamma$) plane.  (a) $\alpha_{1,2}$=0, (b) $\alpha_1=\alpha_2$, (c) $\alpha_1=1.0, \alpha_2=0.0$, and 
	(d)  $\alpha_2=1.0, \alpha_1=0.0$.  Incoherent state and standing wave are denoted by
		IC and SW, respectively. Phase space with bistability (grey shaded region) between  IC and SW  is denoted as R1.
		The Hopf bifurcation (line connected by filled squares) curves are the analytical stability curves.
		Homoclinic  (dotted-dashed line) bifurcation  curve is obtained from XPPAUT.
			}
	\label{fig:1a}
\end{figure*}
\begin{figure}[ht]
	\hspace*{-1.2cm}
	\includegraphics[width=10cm]{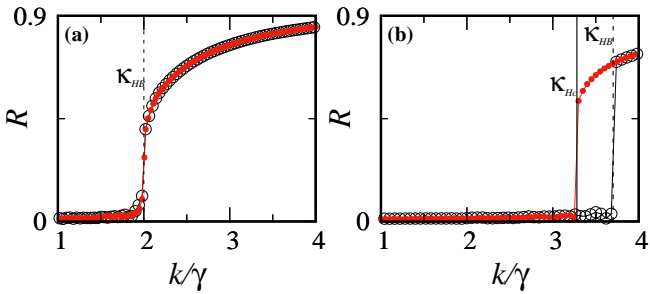}
	\caption{Order parameter (R), obtained from numerical analysis of Eq.~(2) for unimodal distribution, illustrating the nature of the 
	dynamical transitions for (a) $\alpha_{1,2}=0$; (b) $\alpha_1=1, \alpha_2=0.0$.
	}
	\label{fig:1t}
\end{figure}
\begin{figure}[!ht]
	\hspace*{-1cm}
	\includegraphics[width=10cm]{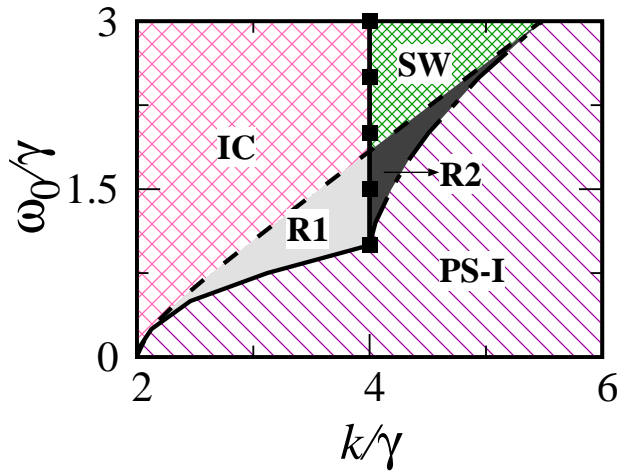}
	\caption{Phase diagram in ($k/\gamma-\omega_0/\gamma$) plane for $\alpha_1=\alpha_2=0$. Incoherent state, partially synchronized state-I and standing wave are denoted by
	 IC,  PS-I and SW, respectively. Phase space with bistability between  IC and PS-I states is denoted as R1 and that between SW and PS-I is denoted as R2.
		The pitchfork (solid black), Hopf bifurcation (line connected by filled squares) and  saddle-node (dashed line) curves are the analytical stability curves.
		Homoclinic  (dotted-dashed line) bifurcation  curve is obtained from XPPAUT.}
	\label{fig:1}
\end{figure}
\begin{figure*}[ht]
	\hspace*{-1.2cm}
	\includegraphics[width=18.5cm]{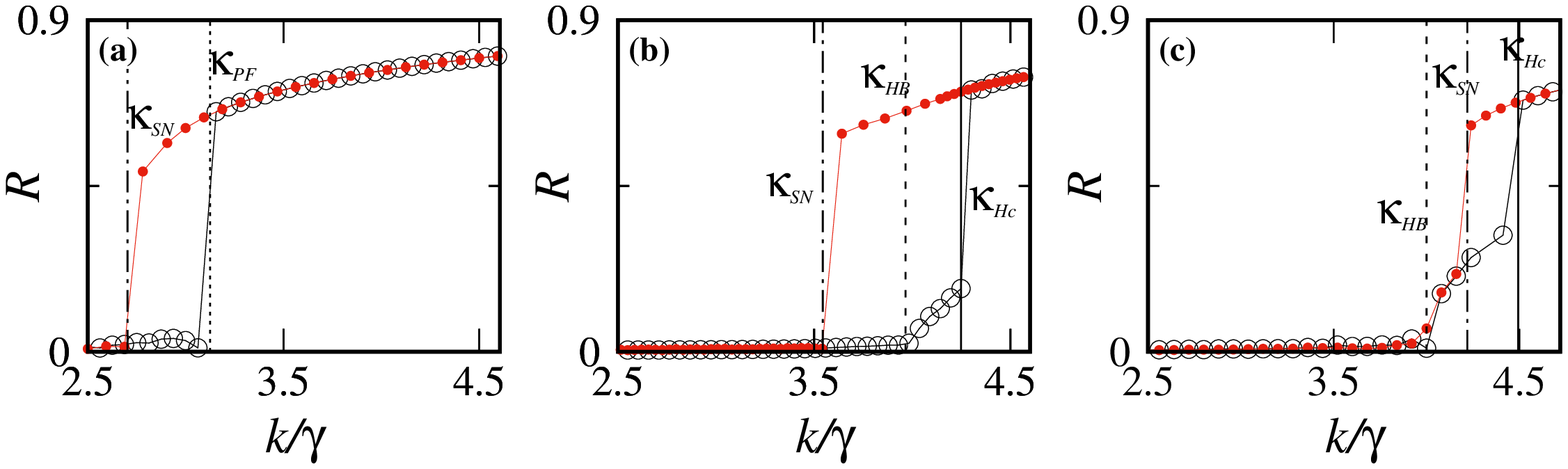}
	\caption{Order parameter (R), obtained from numerical analysis of (2) with bimodal distribution, illustrating the nature of dynamical transitions for $\alpha_{1,2}=0$ for various ratios of $\omega_0/\gamma$: (a) 0.75; (b) 1.5, and (c) 2.
	}
	\label{fig:2t}
\end{figure*}
\subsection{Unimodel Frequency Distribution}
The arbitrary function $a(\omega,t)$ is assumed to satisfy
$|a(\omega,t)| < 1$, together with the requirements that
$a(\omega,t)$ may be analytically
continued in the whole of the complex-$\omega$ plane and it has no singularities in the lower-half complex-$\omega$ plane. Further, $|a(\omega,t)| \to 0$ as 
${\rm Im}(\omega) \to -\infty$. If these conditions are satisfied for $a(\omega,0)$, then as shown in (\ref{eq:OA}), they continue to be satisfied by $a(\omega,t)$ 
as it evolves under Eqs. (\ref{eq:12}) and (\ref{eq:13}). Expanding the unimodal frequency distribution $g(\omega)$, Eq. (\ref{eq:lor}),  in partial fractions as
\begin{align}
g(\omega)=\frac{1}{4\pi i}\bigg[&\frac{1}{((\omega-\omega_0)-i\gamma)}-\frac{1}{((\omega-\omega_0)+i\gamma)}\bigg],
\end{align}
 and evaluating Eq. (\ref{eq:13}) using the appropriate contour integral, the order parameter becomes,
\begin{equation}
Z(t)=a^\star(\omega_0-i\gamma,t).
\end{equation}
Substituting the above in Eq. (\ref{eq:12}), one obtains a
complex ODE, describing the evolution of  the suborder parameter,
\begin{align}
	\frac{\partial Z}{\partial t}+(\gamma+i&\omega_0)Z+\frac{k}{2} Z\bigg[(|Z|^2e^{-i\alpha_{1}}-e^{i\alpha_{1}})\nonumber\\&+|Z|^2 (|Z|^2 e^(-i\alpha_2)-e^{i\alpha_2})\bigg]
	\label{eq:z-dynamics}
\end{align}
Rewriting the above equation in terms of $r$ and $\psi$ as $Z=re^{i\psi}$, one obtains the evolution equations for  $r$ and $\psi$ as
\begin{align}
\dot{r}=&-\gamma r-\frac{k}{2}r((r^2-1)\cos(\alpha_1)  +r^2\cos(\alpha_2)(r^2-1) )\nonumber\\
\dot{\psi}=&-\omega_0-\frac{k}{2}((r^2+1)sin(\alpha_1)+r^2sin(\alpha_2)(r^2+1))
\label{uni}
\end{align}
The above reduced low-dimensional equations describe the dynamics of the model  (\ref{eq:km2})  with unimodal frequency distribution.   Then, $r(t)=\vert \frac{1}{N}\sum_{j=1}^Ne^{\im\theta_j(t)} \vert$  takes either a null value, when the  dynamics corresponds to the incoherent state, or oscillating values corresponding to the standing wave behavior 
of the Sakaguchi-Kuramoto oscillators.

\subsection{Bimodal Frequency Distribution}
Now, we  will deduce the governing equations for the  macroscopic variables for the
 model (\ref{eq:km2}) corresponding to the  bimodal frequency distribution.
 Expanding the bimodal frequency distribution $g(\omega)$, Eq.~(\ref{eq:bil}),  in partial fractions as
\begin{align}
g(\omega)=\frac{1}{4\pi i}\bigg[&\frac{1}{((\omega-\omega_0)-i\gamma)}-\frac{1}{((\omega-\omega_0)+i\gamma)}\nonumber\\&+\frac{1}{((\omega+\omega_0)-i\gamma)}-\frac{1}{((\omega+\omega_0)+i\gamma)}\bigg],
\end{align}
 and evaluating Eq. (\ref{eq:13}) using the appropriate contour integral, the order parameter becomes,
\begin{equation}
Z(t)=\frac{1}{2}[z_{_1}(t)+z_{_2}(t)],
\end{equation}
where
\begin{equation}
z_{_{1,2}}(t)=a^\star(\pm\omega_0-i\gamma,t).
\end{equation}
Substituting the above in Eq. (\ref{eq:12}), one obtains two
coupled complex ODEs, describing the evolution of two suborder parameters,
\begin{align}
\dot{z}_1=&-(\gamma+i\omega_0)z_1+\frac{k}{4}\bigg((z_1+z_2)e^{-i\alpha_1}-{z_1^2}(z_1^\star+z_2^\star) e^{i\alpha_1}\nonumber\\&+\frac{|z_1+z_2|^2}{4}\big( (z_1+z_2)e^{-i\alpha_2}-{z_1^2}(z_1^\star+z_2^\star) e^{i\alpha_2}\big)\bigg),\label{eq:z1}\\
\dot{z}_2=&-(\gamma-i\omega_0)z_2+\frac{k}{4}\bigg((z_1+z_2)e^{-i\alpha_1}-{z_2^2}(z_1^\star+z_2^\star) e^{i\alpha_1}\nonumber\\&+\frac{|z_1+z_2|^2}{4}\big( (z_1+z_2)e^{-i\alpha_2}-{z_2^2}(z_1^\star+z_2^\star) e^{i\alpha_2}\big)\bigg),\label{eq:z2}
\end{align}
where overdot represents the time derivative. 
 Rewriting Eq.~(\ref{eq:z1})~and ~(\ref{eq:z2}) in terms of $r_{_{1,2}}$ and $\psi_{_{1,2}}$, as $z_{_1,2}=r_{_{1,2}}e^{-\im{\psi_{1,2}}}$ and defining the phase difference as $\psi$ = $\psi_{{_1}}-\psi_{{_2}}$ , the dimensionality can be further  reduced to
 three as follows:
 \begin{widetext}
 \begin{subequations}
\begin{align}
\dot{r}_1&=-\gamma r_{_1} - \frac{k}{16}(r_{_1}^2-1)((r_1^2+r_2^2+2 r_1 r_2 \cos[\psi]) (\cos(\alpha_2) r_1+ \cos(\psi+\alpha_2) r_2)+4(\cos(\alpha_1) r_1+\cos(\psi+\alpha_1)r_2)), \\
\dot{r}_2&=-\gamma r_{_2} - \frac{k}{16}(r_{_2}^2-1)((r_1^2+r_2^2+2 r_1 r_2 \cos[\psi]) (\cos(\psi-\alpha_2) r_1+ \cos(\alpha_2) r_2)+4(\cos(\psi-\alpha_1) r_1+\cos(\alpha_1)r_2)), \\
\dot{\psi}&=-2\omega-\frac{k}{16 r_2}(1+r_2^2)(4 r_2\sin(\alpha_1)-4 r_1 \sin(\psi-\alpha_{1})+(r_2 \sin(\alpha_2)-r_1\sin(\psi-\alpha_2))(r_1^2+r_2^2+2 r_1 r_2 \cos[\psi]))\nonumber\\&~~~~~~~-\frac{k}{16 r_1}(1+r_1^2)(4 r_2\sin(\psi+\alpha_1)+4 r_1 \sin(\alpha_{1})+(r_2 \sin(\psi+\alpha_2)+r_1\sin(\alpha_2))(r_1^2+r_2^2+2 r_1 r_2 \cos[\psi])).
\end{align}
\label{eq:ps}
 \end{subequations}
\end{widetext}
The above  system of three coupled  nonlinear ordinary differential equations are the  evolution equations for the macroscopic variables of the
 model (\ref{eq:km2}) and  describes its dynamics  faithfully. Note that the partially synchronized states and standing wave patterns of the 
 Sakaguchi-Kuramoto model  (\ref{eq:km2}) correspond to the periodic and quasi-periodic
orbits, respectively,  in  the above reduced model  (that is the system of three coupled ordinary 
differential equations  governing the evolution of the macroscopic order parameters)   for nonzero $\alpha_{1,2}$. 
However,  for the null value of the asymmetry parameters,  the partially synchronized states and standing wave patterns 
 correspond to the steady states and periodic  orbits, respectively.

\begin{figure*}[ht]
	\hspace*{-1.5cm}
	\includegraphics[width=20cm]{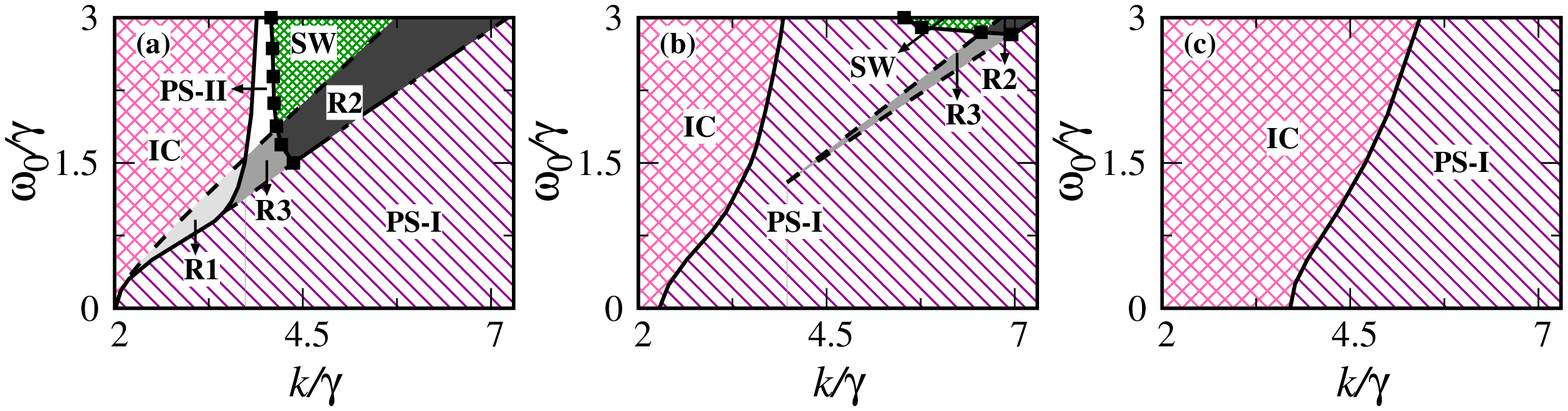}
	\caption{Phase diagrams in  the ($k/\gamma-\omega_0/\gamma$) plane for  $\alpha_1~=~\alpha_2=\alpha$.  (a) $\alpha=0.1$, (b) $\alpha=0.5$,  and (c) $\alpha=1.0$.
Bifurcation curves and dynamical states are represented  similar to those in Fig.~\ref{fig:1}. Partially synchronized state II (PS-II) is observed in the  region enclosed  by pitchfork,
Hopf  and saddle-node bifurcation curves (see Fig.~\ref{fig:3}(a)).  PS-II and R3 in Fig.~\ref{fig:3}(b) are enclosed by saddle-node bifurcation curves.
Here, R3 corresponds to the  region of bistability  between IC, and PS-II states.	}
	\label{fig:3}
\end{figure*}
\section{Phase diagrams of the Sakaguchi-Kuramoto  model with higher order coupling}
\label{sec:numerics}
 In this section, we will proceed to understand the dynamics of the generalized Sagakuchi-Kuramoto model by constructing appropriate two parameter phase 
diagrams and classifying the underlying states from a numerical analysis of the evolution equations of  the macroscopic  order parameters
Eqs.~(\ref{uni}) and (\ref{eq:ps}) corresponding to unimodal and bimodal frequency distributions, respectively.
We also solve the  associated Sakaguchi-Kuramoto  model by numerically integrating Eq.~(\ref{eq:km2}) to verify the dynamical transitions in the phase
diagrams. Specifically,
we will unravel the phase diagrams of the Sakaguchi-Kuramoto  model with higher order coupling and unimodal frequency distribution 
and as well as that with  bimodal frequency distribution for various possible
combinations of the asymmetry parameters.  The number of oscillators is fixed as $N=10^4$ and we use the  standard fourth-order Runge-Kutta integration
scheme with integration step size $h=0.01$ to solve the  Sakaguchi-Kuramoto  model (\ref{eq:km2}).

\subsection{Unimodal Frequency Distribution}
 The reduced low-dimensional equations (\ref{uni}), describing the dynamics of the  Sakaguchi-Kuramoto  model with higher order coupling and unimodal frequency distribution,
 is characterized by a trivial steady state $(r = 0)$, corresponding to the  incoherent state (IC)  and an oscillatory state corresponding to the standing wave (SW) nature of the 
 Sakaguchi-Kuramoto  oscillators. The stability determining eigenvalues of the trivial steady state can be obtained as
\begin{equation}
	\lambda_{1,2}=\frac{-2\gamma+k \cos(\alpha_1)\pm \sqrt{\Delta}}{2},
	\label{eq:M-eigenvalues}
\end{equation}
where
$\Delta=k\sin(\alpha_1)(4\omega_0 +k \sin(\alpha_1))+4\omega_0^2$. The stability condition/curve  for the  onset of IC is 
obtained as 
\begin{align}
	&k_{HB}=2 \gamma  \sec (\alpha_1).
	\label{eq:stability-ISS}
\end{align}

Phase diagrams of the  Sakaguchi-Kuramoto  model with higher order coupling and unimodal frequency distribution for different combinations of the asymmetry parameters
$\alpha_1$ and $\alpha_2$ are depicted in Fig.~\ref{fig:1a}.   The line connected by filled squares corresponds to the  Hopf bifurcation condition (\ref{eq:stability-ISS}). 
In the absence of both the asymmetry parameters, that is for $\alpha_1=\alpha_2=0$,  there is a transition
from the incoherent state to the standing wave pattern as a function of $k$ (see Fig.~\ref{fig:1a}(a)) via the Hopf bifurcation curve. Similar dynamical transition
is also observed for the other choices of the asymmetry parameters, namely for $\alpha_1=\alpha_2=1.0$ (see Fig.~\ref{fig:1a}(b)) and
for $\alpha_1=0$ and $\alpha_2=1.0$  (see Fig.~\ref{fig:1a}(d)) except for the region shift.  For $\alpha_1=1.0$ and $\alpha_2=0.0$, one can observe 
bistability between the IC and SW (indicated by grey shaded region, marked as R1) in Fig.~\ref{fig:1a}(c).  The bistable region is bounded by the 
homoclinic (indicated by dotted-dashed line) and Hopf  bifurcation curves.  Note that the homoclinic bifurcation curve is obtained from XPPAUT.
It is also to be noted that  the dynamical transition  is independent of $\omega_0$ in this case of unimodal frequency distribution, in general.

Now, the time-averaged order paramter $R=\lim_{T\rightarrow\infty}\frac{1}{T}\int_{0}^{T}dt'r(t')$ estimated from the simulation of
the  Sakaguchi-Kuramoto  model, by numerically integrating  Eq.~(\ref{eq:km2}), for the unimodal frequency distribution is
depicted in Fig.~\ref{fig:1t} for two different values of the asymmetry parameters. The line connected by open circles corresponds to the forward trace, 
while the line connected by filled circles corresponds to the backward trace.
 For  $\alpha_1=\alpha_2=0$, there is a 
transition from the  incoherent state (characterized by the  null value of $R$) to the  standing wave pattern, corroborated by a finite value of $R$ (seeFig.~ \ref{fig:1t}(a)),
which is in accordance with the phase diagram in Fig.~\ref{fig:1a}(a) that is obtained from the reduced low-dimensional systems (\ref{uni}).
The dotted line is the analytical Hopf bifurcation curve $k_{HB}$ across which there is a transition.   Similar dynamical transition will be
observed for the other combinations of the asymmetry parameters except for the region shift as in the  phase diagrams (see Fig.~\ref{fig:1a})
and hence they are not shown here to avoid repetitions.
Nevertheless, there is a bistability between IC and SW as in the phase diagram for $\alpha_1=1.0$ and $\alpha_2=0.0$ (see \ref{fig:1t}(b))
bounded by the homoclinic  and Hopf  bifurcation curves. Thus, direct numerical simulation of the model equation agrees  well with the dynamical
transitions observed from their reduced low-dimensional equations corresponding to the macroscopic order parameters.

\subsection{Bimodal Frequency Distribution}
\subparagraph{Case I ($\alpha_1=\alpha_2$~=~0):}
In order to appreciate and understand the effect of the asymmetry parameters $\alpha_1$ and $\alpha_2$  on the dynamics as represented by
the phase diagram, one should  first familiarize with the
phase diagram of the Sakaguchi-Kuramoto  model with higher order coupling and bimodal frequency distribution in the absence of the asymmetry parameters.
The phase diagram in the ($\omega_0/\gamma$-$k/\gamma$) plane for  the  case $\alpha_1=\alpha_2=0$ is depicted in Fig.~\ref{fig:1}.  The dynamical states
in the phase diagram are distinguished by features which are essentially based on the asymptotic behavior of $r(t)$.  Incoherent state (IC), partially synchronized state (PS-I) and standing wave (SW)
along with the bistable regimes (R1 and R2) among the observed dynamical states are depicted in the phase diagram.  The parameter space marked as R1 corresponds
to the bistable regime between IC and PS-I states, while that indicated as R2 corresponds to the bistable regime between SW and PS-I states. The null value of $r(t)$ 
characterizes the incoherent state, while a finite value of $r(t)$  indicates  partially synchronized states.  Oscillating nature of $r(t)$  
confirms the standing wave. 

The stable regions of the incoherent state in the phase diagram can be inferred from the dynamical equations of the reduced macroscopic variables 
given in Eqs.~(\ref{eq:z1})~and~(\ref{eq:z2}).  The phases of the oscillators are uniformly distributed between $0$ to $2\pi$ for 
the incoherent state and hence it is characterized by $z_1 = z_2 = 0$. Performing a linear stability analysis of the fixed point $(z_1,~z_2)=(0,~0)$,
 one obtains the condition for stability as
	\begin{align}
	 k_{PF}=\frac{2(\gamma^2+\omega_0^2)}{\gamma}, ~~~~\text{for} ~\omega_0/\gamma~~<~1,\\
	 k_{HB}=4\gamma ~~~~~~~~~  \text{for} ~\omega_0/\gamma~~\ge~1.
		\label{eq:pf}
	\end{align}
Here, $K_{PF}$ corresponds to the  pitchfork bifurcation curve across which  the fixed point $(z_1,~z_2)=(0,~0)$ (incoherent state) loses
 its stability leading to  the inhomogeneous steady state (PS-I state), while $K_{HB}$ corresponds to the  Hopf bifurcation curve  across
 which the incoherent state loses its stability  resulting in the standing wave pattern. The pitchfork bifurcation curve, 
 indicated by the solid line in Fig.~\ref{fig:1}, serves as the
 boundary between the incoherent and partially synchronized state
for   $\omega_0/\gamma~~<~1$. The Hopf bifurcation curve, denoted by the line connected by filled squares,  demarcates the incoherent state and standing wave region
of the phase diagram.  The dashed line in Fig.~\ref{fig:1} corresponds to the saddle-node bifurcation curve, while  the homoclinic bifurcation curve is denoted as the  dotted-dashed line.
The latter  is obtained from the software XPPAUT~\cite{xpp}, while the former is determined as follows.
The inhomogeneous steady state of the PS-I region in the phase diagram is characterized by $r_1=r_2=r=Const.$ and $\psi_1=-\psi_2=\phi$,
and hence from  Eqs.~(\ref{eq:ps}) one can obtain
\begin{subequations}
\begin{align}
	\sin(2\phi)&=\frac{8\omega_0}{k(1+r^2)(2+r^2+r^2\cos(2\phi))},\\
	\cos(2\phi)&=\frac{k-kr^2-a}{(k r^2(r^2-1))},
\end{align}
\end{subequations}
 where $a$~=$\sqrt{k^2-2k^2r^2+8k\gamma r^2+k^2r^4-8k\gamma r^4}$. The above equations give the following solutions for the stationary $r$ and $\phi$:
 \begin{subequations}
 \begin{align}
 	1&=\frac{64 \omega_0^2(r^2-1)^2}{(r^2+1)^2(k-k r^2+ a)^2}+\frac{(k r^4-k+a)^2}{(k^2 r^4(r^2-1)^2)}\\
 	\tan(2\phi)&=\frac{8 k \omega_0 r^2 (r^2-1)^2}{(r^2+1)(k-k r^2+a)(k r^4-k+a)}.
 \end{align}
 \label{sn}
 \end{subequations}
 Now, one can  numerically solve the above equations for fixed values
of the parameters to obtain $r$ and $\phi$, which can be substituted back in the original equation of motion of the order parameters, Eqs.~(\ref{eq:ps}),
to deduce the characteristic eigenvalue equation. The resulting eigenvalues determine the saddle-node bifurcation curves 
in the ($\omega_0/\gamma$-$k/\gamma$)  parameter space.

The standing wave  pattern loses its stability across the  homoclinic bifurcation curve resulting in the PS-I state.
Upon decreasing the value of  $k/\gamma$ in the phase diagram, the PS-I state (inhomogeneous steady states of $z_1$ and $z_2$) loses its stability via  the saddle-node 
bifurcation curve resulting in the incoherent state ($z_1 = z_2 = 0$) up to $\omega_0/\gamma=1.6$  and  in the standing wave patterns
for  $\omega_0/\gamma>1.6$.   Hence, the bistability between the IC and PS-I states is enclosed by the saddle-node and pitchfork bifurcation curves in the phase diagram
in the region denoted as R1. Saddle-node and homoclinic bifurcation curves enclose the bistable region between the standing wave and PS-I state, which is denoted as R2 in the 
phase diagram. It is to be noted that the phase diagram of the Sakaguchi-Kuramoto  model with higher order coupling and bimodal frequency distribution 
in the absence of asymmetry parameters resembles closely that of the Sakaguchi-Kuramoto  model with  pairwise interactions and bimodal frequency distribution~\cite{bim}.  The 
higher order coupling  has essentially enlarged the bistable regions of the phase diagram.   Further, the Sakaguchi-Kuramoto  model with higher order coupling and bimodal frequency distribution is characterized by PS-I, R1 and R2 when compared to  the Sakaguchi-Kuramoto  model with higher order coupling and unimodal frequency distribution 
(compare Figs.~\ref{fig:1}  and ~ \ref{fig:1t}(a)). Similar rich dynamical states are also observed for the other choices of the asymmetry parameters in the
presence of bimodal frequency distribution as will be elucidated in the following cases.

 Now, the order parameter $R$ estimated from
the  Sakaguchi-Kuramoto  model by numerically integrating  Eq.~(\ref{eq:km2}) for the bimodal frequency distribution is
depicted in Fig.~\ref{fig:2t} for the asymmetry parameters $\alpha_1=\alpha_2=0$ and for three different values of $\omega_0/\gamma$. 
Here, the line connected by open circles corresponds to the forward trace, while the line connected by filled circles corresponds to the backward trace 
as in Fig.~\ref{fig:1t}.  The dotted vertical line  in Fig.~\ref{fig:2t}(a) corresponds to the  analytical pitch-fork bifurcation curve,
the dotted-dashed line corresponds to the analytical saddle-node bifurcation curve,  and the dashed line in Fig.~\ref{fig:2t}(b) corresponds to the analytical
Hopf bifurcation curve, while the solid line corresponds to the homoclinic bifurcation curve obtained using XPPAUT.
There is a transition from the incoherent state to the standing wave via the  pitch-fork bifurcation during the forward trace, whereas there is
a transition from the SW to IC via the saddle-node bifurcation during the reverse trace  (see Fig.~\ref{fig:2t}(a)) for $\omega_0/\gamma=0.75$.
Similarly, there is a transition from
IC(SW) to SW(IC) via  the homoclinic(saddle-node)  bifurcation curve during the forward(backward) trace for  $\omega_0/\gamma=1.5$
as depicted in Fig.~\ref{fig:2t}(b).   For  $\omega_0/\gamma=2.0$, there is a similar transitions via the homoclinic and saddle-node bifurcation curves
during the forward and  backward traces, respectively. These transitions, obtained by numerically solving the model equation  (\ref{eq:km2}),  
perfectly correlate with the dynamical transitions observed in the phase diagram (see Fig.~\ref{fig:1}), which are obtained by solving the 
reduced low-dimensional evolution equations for the macroscopic order parameters (\ref{eq:ps}).

 \begin{figure*}[ht!]
	\hspace*{-1cm}
	\includegraphics[width=19.2cm]{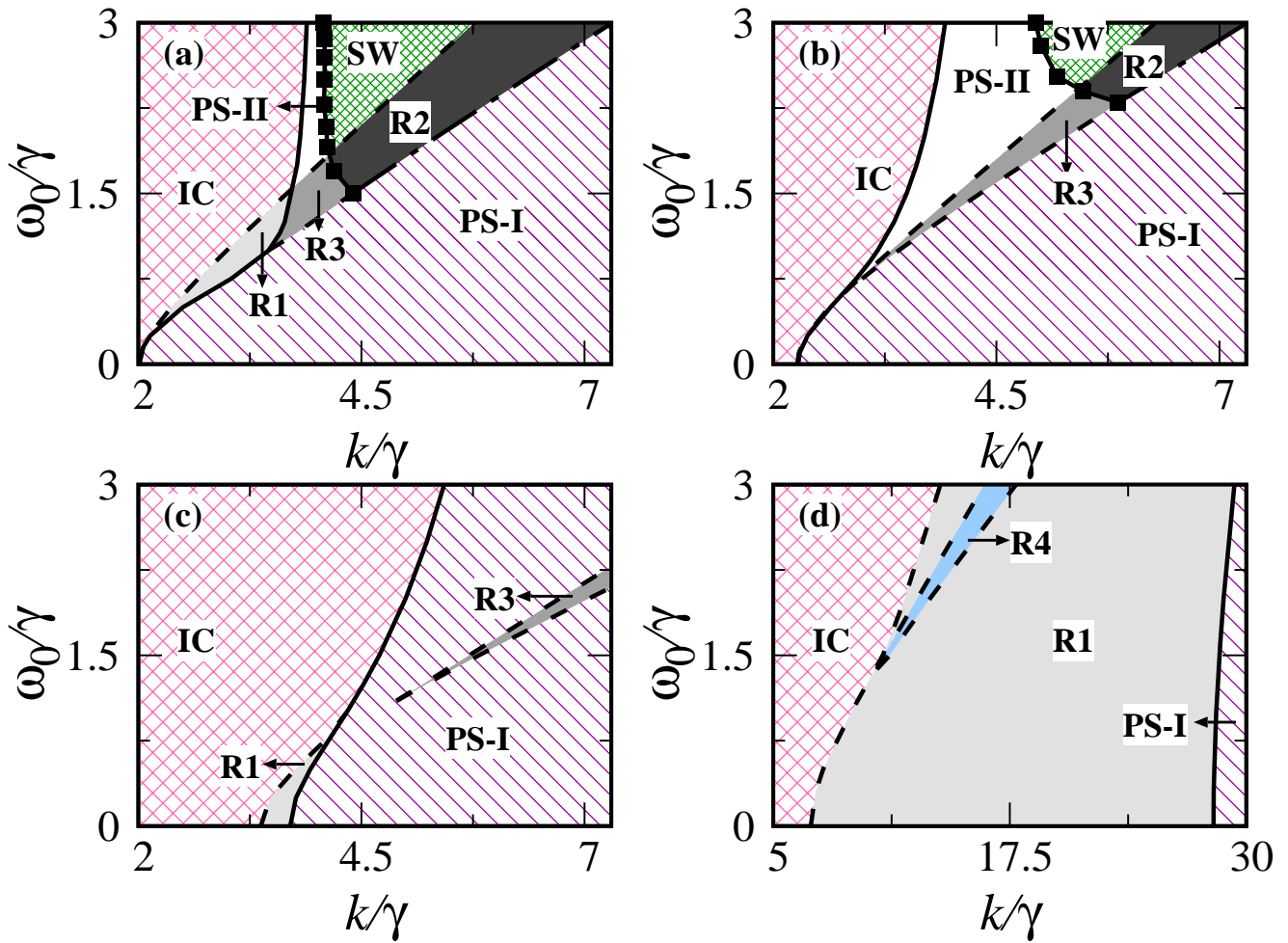}
	\caption{ Phase diagrams in  the ($k/\gamma-\omega_0/\gamma$)  plane for  $\alpha_2~=~0$ and for various values of  the asymmetry parameter in the
	pairwise coupling.  (a) $\alpha_1=0.1$, (b) $\alpha_1=0.5$, (c) $\alpha_1=1.0$ and (d) $\alpha_1=1.5$. Bifurcation curves and dynamical states are similar to those in
	Fig.~\ref{fig:3}(a). Here, R4 corresponds to the  region of multistability  between IC, PS-I and PS-II states.}
	\label{fig:4}
\end{figure*}
\subparagraph{Case II ($\alpha_1=\alpha_2~\ne$~0):}
In order to analyze the effect of the asymmetry parameters on the phase diagram (see  Fig.~\ref{fig:1}),  we  have next  considered  the case where the asymmetry parameters
 $\alpha_1~=~\alpha_2~=~\alpha$ for simplicity.
We have depicted the corresponding phase diagrams in the  ($k/\gamma-\omega_0/\gamma$)  plane in Figs.~\ref{fig:3}(a)-\ref{fig:3}(c) for $\alpha=0.1, 0.5$,  and $1$, respectively.  
The dynamical sates
and the bifurcation curves are similar to those in Fig.~\ref{fig:1} without any asymmetry parameter.  However for  $\alpha=0.1$,  partially synchronized state-II (PS-II)
is characterized by a different set of inhomogeneous steady states corresponding to nonzero values of  ($z_1,~z_2$)   in addition to the dynamical states observed in Fig.~\ref{fig:1}. 
 A linear stability analysis of the fixed point $(z_1,~z_2)=(0,~0)$  results in the stability condition
\begin{align}
\hspace{-0.5cm}
\omega_0^2 = \frac{{(32 \gamma^3 k + 
\gamma k^3) \cos(\alpha_1) -32 \gamma^4 - 6 \gamma^2 k^2 - 4 \gamma^2 k^2 \cos(2 \alpha_1)}}{{
32 \gamma^2 + k^2 - 16 \gamma k \cos(\alpha_1) + k^2 \cos(2 \alpha_1])}}.
	\end{align}
The above algebraic expression can be further simplified as
	\begin{align}
	 \gamma &k^3 \cos(\alpha_1)-32 \gamma^4 - 6 \gamma^2 k^2 + 32 \gamma^3 k \cos(\alpha_1)\nonumber\\&- 
		2 \omega_0^2 ( k \cos(\alpha_1)-4 \gamma )^2 - 4 \gamma^2 k^2 \cos(2 \alpha_1)=0,
		\label{eq:pf1}
	\end{align}
which actually corresponds to the  pitchfork bifurcation curve across which  the fixed point $(z_1,~z_2)=(0,~0)$ (incoherent state) loses
 its stability leading to  the partially synchronized states  PS-I and PS-II. Note that the incoherent state loses it stability only through the pitchfork bifurcation curve 
in the entire explored range of   $\omega_0/\gamma$ (see Fig.~\ref{fig:3}(a)).
 All other bifurcation curves are obtained from XPPAUT.  One may observe that the PS-II state
is enclosed by pitchfork, Hopf and homoclinic bifurcation curves, whereas the region corresponding to the bistability  between  PS-I and PS-II (denoted by R3)
is enclosed by pitchfork, Hopf and  saddle-node bifurcation curves. The other dynamical transition and bistable regions are similar to that discussed in  
Fig.~\ref{fig:1} in the absence of the asymmetry parameters.  
Thus, a rather low value of the asymmetry parameters results in an additional partially synchronized state (PS-II state) with a region
of multistability between PS-I and PS-II.  

However,  a slight increase in the values of the asymmetry parameters results in drastic changes in the phase diagram (see Fig.~\ref{fig:3}(b) for $\alpha=0.5$).  It is evident
from the figure that the bistable regions (R2 and R3) and the parameter space with standing wave are reduced drastically with increase in the PS-I state.  
The PS-II state coexists with the PS-I state in the region enclosed by the two saddle-node bifurcation curves, while the  bistable region R1 is completely wiped off from the phase diagram.
 A large asymmetry parameter results in the loss of bistable regions and standing wave regions completely from the phase diagram, while retaining only
the incoherent state and partially synchronized state-I  as illustrated in Fig.~\ref{fig:3}(c) for $\alpha=1$.   Further increase in the asymmetry parameter results in 
 similar phase diagrams as in  Fig.~\ref{fig:3}(c).

\begin{figure*}[ht!]
     \hspace*{-1cm}
    \centering
	\includegraphics[width=19.5cm]{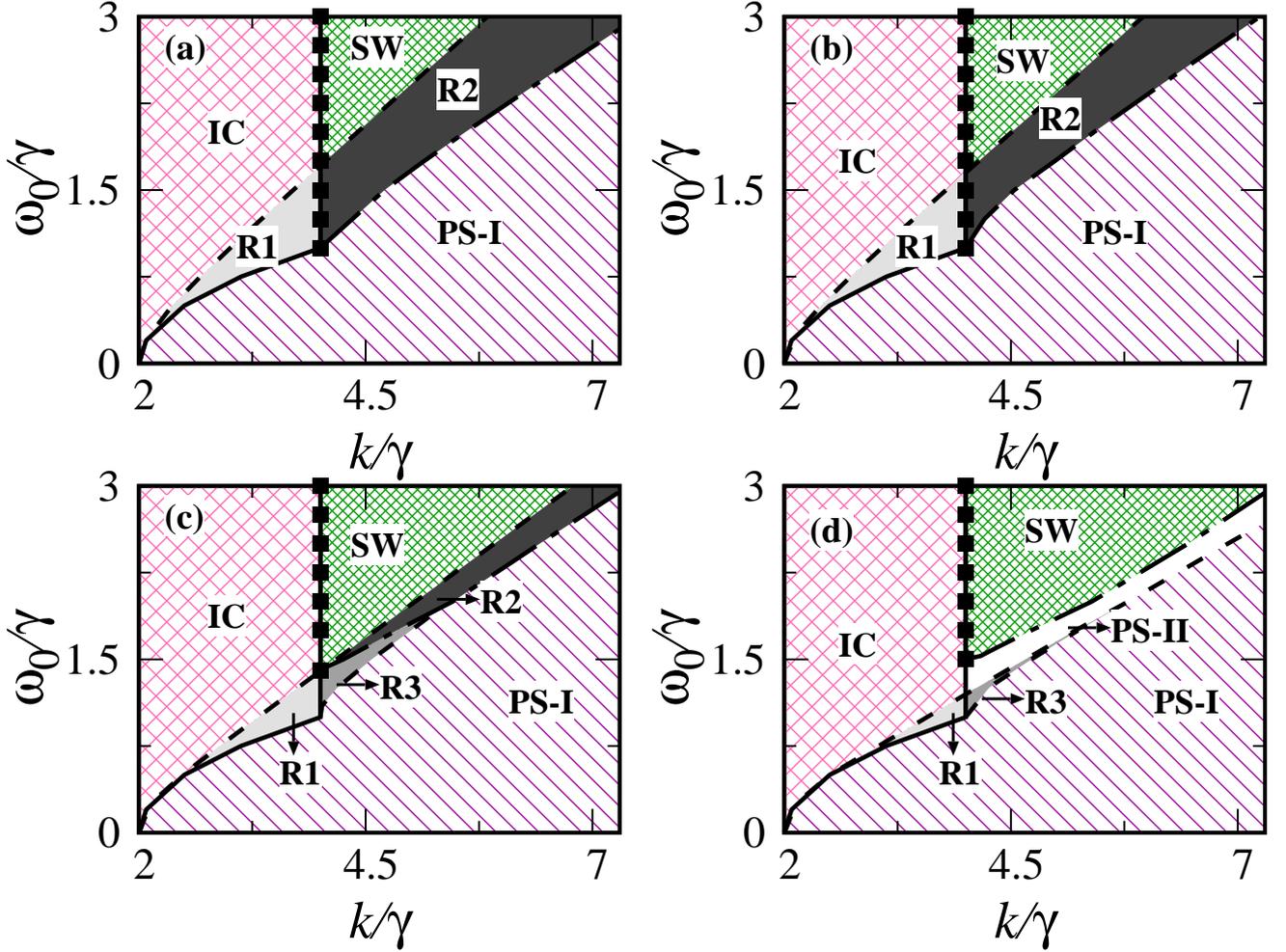}
        \caption{Phase diagrams in  the ($k/\gamma-\omega_0/\gamma$)  plane for  $\alpha_1~=~0$ and for various values of  the asymmetry parameter in the
	higher order coupling.  (a) $\alpha_2=0.1$, (b) $\alpha_2=0.5$, (c) $\alpha_2=1.0$ and (d) $\alpha_2=1.5$. Bifurcation curves and dynamical states are similar to those in
	Fig.~\ref{fig:3}(a). }
        	\label{fig:2}
\end{figure*}
\subparagraph{Case III  ($\alpha_1~\ne$~~0; $\alpha_2$~=~~0): }
Now, we analyse the nature of the phase diagram with asymmetry parameter only in the pairwise coupling by switching off the asymmetry parameter in the
higher order coupling, so that $\alpha_1~\ne~~0$ and  $\alpha_2~=~~0$.  The phase diagrams for $\alpha_1=0.1, 0.5, 1$  and $1.5$
 are shown in Figs.~\ref{fig:4}(a)-\ref{fig:4}(d), respectively. 
For $\alpha_1=0.1$, the dynamics and the dynamical transitions in the phase diagram (see Fig.~\ref{fig:4}(a)) are similar to those observed in Fig.~\ref{fig:3}(a) for
 $\alpha_1~=~\alpha_2=0.1$, which elucidates that the onset of  PS-II state is facilitated by the asymmetry parameter in the pairwise coupling and is  independent of the
 asymmetry parameter in the higher order coupling.   Increasing $\alpha_1$ to $\alpha_1=0.5$  results in an enhancement of the PS-II state 
 and the bistability between both the partially synchronized states in the phase diagram (see Fig.~\ref{fig:4}(b)). It is to be noted that R3 is enclosed by the 
 saddle-node and Hopf bifurcation curves.  The  spread of SW and R2  in the phase diagram is decreased appreciably for increasing values of $\alpha_1$, whereas
 that of PS-I remains almost unaffected.  The bistability between the IC and PS-I (region R1)  states is completely destroyed.

Next, the phase diagram for $\alpha=1$ is depicted in Fig.~\ref{fig:4}(c), where the spread of  SW and R2  is completely eliminated. 
Further, the spread of R3 enclosed by the saddle-node bifurcation curves in the phase diagram is considerably reduced.
It is to be noted that  there is a reemergence of the bistable region R1 even for $\omega_0/\gamma~<~1/\sqrt{3}$, where bimodal frequency distribution 
becomes unimodal, which elucidates that the bistable region R1 has its manifestation in the phase diagram essentially due to  the higher order coupling. 
Otherwise, the phase diagram is almost equally shared by IC and PS-I states.  Further increase in the asymmetry parameter in the pairwise coupling
results in the increase in the R1 region to a large extent, where IC and PS-I  states coexist and are bounded by the saddle-node and pitchfork bifurcation curves.   It is to
be noted that a new multistable region enclosed by the saddle-node bifurcation curves appears (denoted as R4 in  Fig.~\ref{fig:4}(d)  for $\alpha=1.5$), where IC, PS-I and PS-II states
coexist.  Thus, it is evident that  the asymmetry parameter in the pairwise coupling facilitates several interesting multistable states in the phase diagram mediated by
various types of bifurcations.

\begin{figure*}[ht!]
     \hspace*{-1cm}
    \centering
	\includegraphics[width=19.5cm]{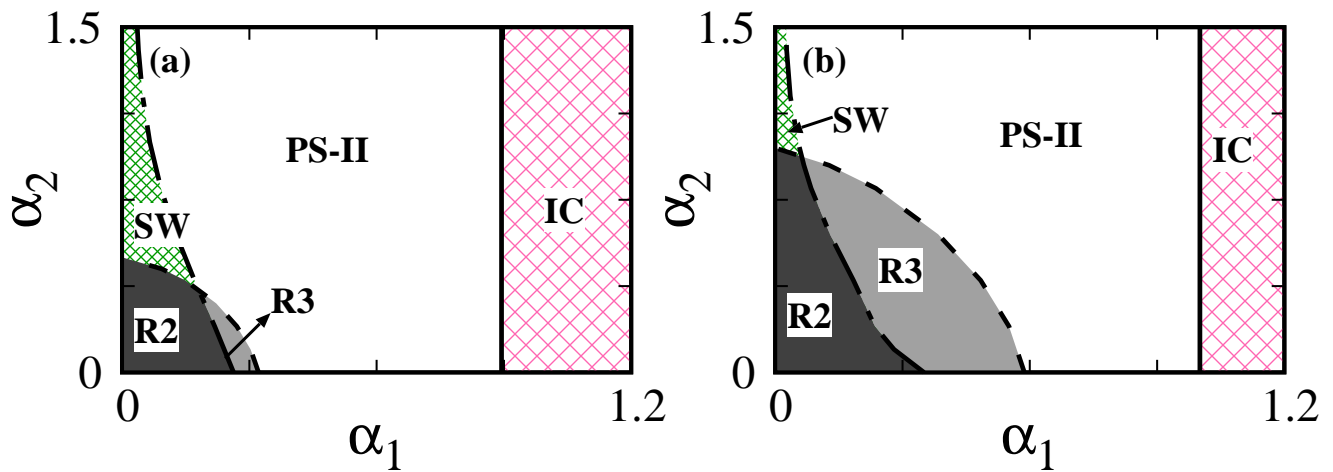}
        \caption{Phase diagrams in  the ($\alpha_1-\alpha_2$)  plane for  $\omega_0/\gamma~=~2$.
	 (a) $k/\gamma=4.5$, and (b) $k/\gamma=5.0$. Bifurcation curves and dynamical states are similar to those in
	Fig.~\ref{fig:3}(a). }
        	\label{fig:5}
\end{figure*}
\subparagraph{Case IV  ($\alpha_2~\ne$~~0; $\alpha_1$~=~~0): }
In order to analyze the effect of the asymmetry parameter in higher order interactions alone, we  have fixed $\alpha_1=0$ and depicted the phase diagrams
in Figs.~\ref{fig:2}(a)-\ref{fig:2}(d) for $\alpha_2=0.1, 0.5, 1.0$ and $1.5$, respectively.  The phase diagram (see Fig.~\ref{fig:2}(a)) for $\alpha_2=0.1$ is similar to
the phase diagram in Fig.~\ref{fig:1}, which is depicted for  the choice $\alpha_1=\alpha_2=0$,  but  now with an enlarged  bistable region R2 enclosed by saddle-node and homoclinic 
bifurcation curves.  Thus, it is again evident  that
the asymmetry parameters largely contribute to the onset of multistability and facilitate the latter to a large extent.  Note that the  PS-II  state  and consequently the region R3 are
absent in the phase diagram for $\alpha_1=0$, which is actually facilitated by intermediate values of $\alpha_1$  (see Figs.~\ref{fig:4}(a) and ~\ref{fig:4}(b)).
Increasing $\alpha_2$ to  $0.5$ (see Fig.~\ref{fig:2}(b)), the spread of the  bistability region shrinks  compared to that in  Fig.~\ref{fig:2}(a).  Further increase
in the value of the asymmetry parameter in the higher order coupling results in a decrease in the spread of R2  with the onset of R3, where PS-I and PS-II coexist,
via  the saddle-node bifurcation as depicted in Fig.~\ref{fig:2}(c) for $\alpha=1.0$.  For further larger values of $\alpha_2$, the spread of R1  and R3 in the phase
diagram decreases to a large extent  resulting in the monostable regions of IC, PS-I, PS-II and SW states as depicted in Fig.~\ref{fig:2}(d) for $\alpha=1.5$. 
The spread of R2 is completely wiped off from the phase diagram  for  $\alpha_2=1.5$. Thus, it is evident that large values of $\alpha_2$ facilitate the onset of
PS-II and eventually R3, while smaller values of  $\alpha_2$ favor the spread of bistable regions to a large extent.

\subparagraph{Case V  ($\alpha_1~>0~\&~\alpha_2>0$): }
Now, we consider $\alpha_1>0$ and  $\alpha_2>0$ in order to analyze the dynamical states and their transitions due to the trade-off between the asymmetry parameters
 in both pair-wise and higher order couplings. Phase diagrams in the asymmetry parameter  ($\alpha_1, \alpha_2$)  space for $\omega_0/\gamma=2$ and 
  for two different values of $k/\gamma$ are depicted in Figs.~\ref{fig:5}.  The dynamical states and their bifurcation transitions are found to  be similar to those
  in the previous figures.  For low values of $\alpha_2$, there is a transition from  R2 to R3  via the homoclinic bifurcation and then to PS-II state via the saddle-node bifurcation and
  finally to IC state  through  the pitchfork bifurcation as a function of $\alpha_1$.
  For larger values of  $\alpha_2$, there is a transition from SW to PS-II via the homoclinic bifurcation  and then to IC via  the pitchfork bifurcation as 
  a function of $\alpha_1$. Small  to intermediate values of $\alpha_1$  and $\alpha_2$ favor bistable states R2 and R3, while larger values of the asymmetry parameters
   $\alpha_1$  and/or $\alpha_2$ result in monostable states (see Figs.~\ref{fig:5} and ~\ref{fig:3}). Increasing $k/\gamma$ from $4.5$ to $5$ results in increase in the spread of bistable regions R2 and R3 (compare
  Figs.~\ref{fig:5}(a) and ~\ref{fig:5}(b) ).

\section{Summary and Conclusion}
\label{sec:conclusions}
Higher order interactions have physical relevance in physics and neuroscience and they have gained recent interest in network theory.  In this work, we have investigated the 
phase diagrams of the Sakaguchi-Kuramoto model along with a  higher order interaction, and unimodal and bimodal distributions of the natural frequencies of the  individual phase oscillators.
We  have also introduced asymmetry parameters  both in the pairwise and higher order couplings to elucidate their role in the dynamical transitions in the phase diagram.
We have investigated the effects of five possible combinations of the asymmetry parameters $\alpha_1$ and $\alpha_2$ on the phase diagram along with  the higher order interaction.
Using the Ott-Antonsen ansatz, we have obtained the coupled evolution equations corresponding to the macroscopic order parameters.
We have deduced the analytical stability condition for the linear stability of the incoherent state, resulting in  the pitchfork bifurcation curve,  using the governing equations of the  
macroscopic order parameters.  Further, we have also analytically deduced the Hopf bifurcation curve for $\alpha_1=0$, while the saddle-node and homoclinic bifurcation
curves are obtained using the software package  XPPAUT. 
The Sakaguchi-Kuramoto model along with a higher order interaction and  unimodal frequency distribution displays only IC and SW states,
and bistability among them  for $\alpha_1=1.0$ and $\alpha_2=0.0$.  In contrast,
we have observed rich phase diagrams with dynamical states such as IC, PS-I, PS-II and SW states along with the
bistable (R1, R2 and R3) and multistable (R4) states with the bimodal frequency distribution.

In the absence of asymmetry parameters, higher order couplings favor the spread of the bistable states R1 and R2 to a large extent when compared to the
Sakaguchi-Kuramoto model  with pairwise coupling alone and  bimodal frequency distribution.  Further, the asymmetry parameters favor the onset of the
bistable regions R3 and R4 which are generally absent in the Sakaguchi-Kuramoto model  with pairwise coupling and  bimodal frequency distribution.  It is to be
noted that rather low values of the asymmetry parameter in the pairwise coupling for $\alpha_2=0$ and
relatively larger values of  the asymmetry parameter in the higher order  coupling  for $\alpha_1=0$ favors the onset of  PS-II state  and eventually the region R3 in the phase diagrams.  
However, very large values of both the asymmetry parameters render the phase diagram only with monostable dynamical states.  
It is to be noted that there exists  bistable region R1 even for $\omega_0/\gamma~<~1/\sqrt{3}$ in the phase diagrams, where the bimodal frequency distribution 
breaks down to unimodal one, which is purely a manifestation of the higher order coupling as the bistable region R1, which has not yet been observed in the
Sakaguchi-Kuramoto model with pairwise coupling only along with unimodal frequency distribution.  We sincerely believe that the above results, with rich phase diagrams
comprising of bistable and monostable regions of
the Sakaguchi-Kuramoto model  due to the tradeoff between the asymmetry parameters and the higher order coupling, provide valuable new insights on
the dynamical nature of the model.  Note that the presence of bistable (multistable) regions denote the regions across which abrupt dynamical transition occurs,
a typical nature of biological systems and, in particular, in  neuroscience where bistability and fast switching between states are very relevant.
 

\section*{Acknowledgements}
 M.M. thanks the Department of
Science and Technology, Government of India, for providing financial support through an INSPIRE Fellowship No.
DST/INSPIRE Fellowship/2019/IF190871. 
DVS  is supported by the DST-SERB-CRG Project under Grant No. CRG/2021/000816.
The work of V.K.C. is supported 
by the SERB-DST-MATRICS Grant No. MTR/2018/000676 and  DST-SERB-CRG Project under Grant No. CRG/2020/004353 and VKC wish to thank DST, New Delhi for computational
facilities under the DST-FIST programme (SR/FST/PS- 1/2020/135) to the Department of Physics.  ML is supported by the DST-SERB National Science Chair program.


\end{document}